\documentclass[sigconf]{acmart}

\usepackage{./lib/_macros}

\usepackage{orcidlink}
\definecolor{orcidlogocol}{HTML}{A6CE39}

\usepackage{xr}
\usepackage{hyperref}
\AtBeginDocument{%
  \providecommand\BibTeX{{%
    \normalfont B\kern-0.5em{\scshape i\kern-0.25em b}\kern-0.8em\TeX}}}

\copyrightyear{2023}
\acmYear{2023}
\setcopyright{acmlicensed}\acmConference[VRST 2023]{29th ACM Symposium on Virtual Reality Software and Technology}{October 9--11, 2023}{Christchurch, New Zealand}
\acmBooktitle{29th ACM Symposium on Virtual Reality Software and Technology (VRST 2023), October 9--11, 2023, Christchurch, New Zealand}
\acmPrice{15.00}
\acmDOI{10.1145/3611659.3615710}
\acmISBN{979-8-4007-0328-7/23/10}

\begin{document}

\title[Pointing Performance on Virtual and Physical Large Curved Displays]{Exploring Users' Pointing Performance on Virtual and Physical Large Curved Displays}

\author{A K M Amanat Ullah}
\affiliation{%
  \institution{University of British Columbia}
  \city{Okanagan}
  \country{BC, Canada}}
\email{amanat7@student.ubc.ca}

\author{William Delamare}
\affiliation{%
  \institution{Univ. Bordeaux, ESTIA-Institute of Technology, Bidart, France}
  \city{}
  \country{}}
\email{william.delamare@acm.org}

\author{Khalad Hasan}
\affiliation{%
  \institution{University of British Columbia}
  \city{Okanagan}
  \country{BC, Canada}}
\email{khalad.hasan@ubc.ca}

\renewcommand{\shortauthors}{Ullah et al.}

\newcommand{\amanat}[1]{{\textcolor{brown}{#1}}} 
\newcommand{\remove}[1]{{\textcolor{pink}{#1}}} 
\newcommand{\update}[1]{{\textcolor{black}{#1}}} 
\begin{abstract}
  Large curved displays have emerged as a powerful platform for collaboration, data visualization, and entertainment. These displays provide highly immersive experiences, a wider field of view, and higher satisfaction levels. Yet, large curved displays are not commonly available due to their high costs. With the recent advancement of Head Mounted Displays (HMDs), large curved displays can be simulated in Virtual Reality (VR) with minimal cost and space requirements. However, to consider the virtual display as an alternative to the physical display, it is necessary to uncover user performance differences (e.g., pointing speed and accuracy) between these two platforms. In this paper, we explored users’ pointing performance on both physical and virtual large curved displays. Specifically, with two studies, we investigate users' performance between the two platforms for standard pointing factors such as target width, target amplitude as well as users' position relative to the screen. Results from user studies reveal no significant difference in pointing performance between the two platforms when users are located at the same position relative to the screen. In addition, we observe users' pointing performance improves when they are located at the center of a semi-circular display compared to off-centered positions. We conclude by outlining design implications for pointing on large curved virtual displays. These findings show that large curved virtual displays are a viable alternative to physical displays for pointing tasks.
 
\end{abstract}








\begin{teaserfigure}
\begin{center}
  \includegraphics[width=0.8\textwidth]{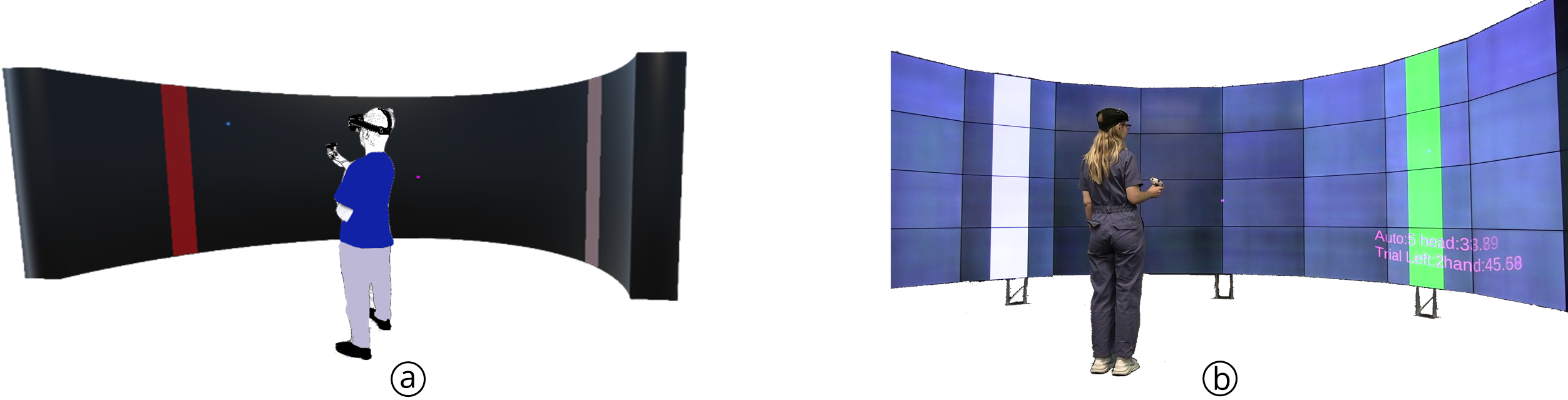}
  \end{center}
  \caption{Pointing on large curved displays: a participant using (a) virtual and (b) physical displays.}
  \Description{Pointing on large curved displays: a participant using (a) virtual and (b) physical displays.}
  \label{fig:teaser}
\end{teaserfigure}

\maketitle

\section{Introduction}

Large displays are widely used for collaboration \cite{andrews2011information, card1999information, baudisch2001focus, cockburn2009review}, data visualization \cite{novak2008designing, satyanarayan2013chi, birnholtz2007exploratory}, and entertainment \cite{ardito2015interaction, scheible2005mobilenin, khoo2009designing}. 
However, hardware and space requirements make it challenging to use large displays \cite{cavallo2019dataspace}.
Due to the advancement in Virtual Reality (VR) technology, large virtual screens are emerging as a viable option to replace physical ones \cite{Ullah2023LargeCurved}. 
Although large physical displays have been explored extensively in previous research, little work has been done on virtual displays - especially comparing users' performance between these two platforms to explore the potential of using large virtual displays as an alternative to physical displays.

Numerous studies have used Fitts law to explore users' pointing performance on large physical displays \cite{tao2021freehand, babic2018pocket6, lischke2016magic, liu2015gunslinger, haque2015myopoint, siddhpuria2018pointing, nancel2013high}. 
These studies revealed that users' performance is significantly affected by target width and amplitude (i.e., the distance between two targets in meters or pixels).
Previous studies \cite{shi2022pointing, hansen2018fitts, batmaz2022effect, batmaz2019head, chowdhury2022wriarm} on target selection in VR evaluated pointing performance using Fitts law in a 3D environment rendered via an HMD.
These studies in VR also concluded that user performance primarily depends on target widths and amplitudes.
In addition, researchers showed that large curved displays offer many advantages over flat displays, as the user can be situated at a uniform distance from the center, offering higher levels of immersion and satisfaction \cite{kyung2021curved, shupp2009shaping}. 
However, there is limited research on pointing performance on curved displays as previous work on curved displays focuses mainly on viewing tasks \cite{ahn2014research, shupp2009shaping, urakami2021comparing} and reading tasks \cite{kyung2021curved, park2017effects}.  
Due to the gap in previous literature on pointing performance on virtual and large curved displays, it is critical to compare users' pointing performance between virtual and physical large curved displays.

\update{We hypothesize that large virtual displays can be an alternative to physical displays for investigating users' pointing-related performance.}  
To validate our hypothesis, we conduct user studies comparing users' performance between the two platforms - large physical and virtual curved displays with the same experimental setup.   
We consider standard pointing factors (e.g., target width and movement amplitude) and curve-related user position (e.g., off-centered position and distance from the center of the screen). 
We performed two user studies.
In the first study, we explore the difference between the physical and the virtual platform on users’ pointing performance with 1D Fitts' law tasks when users are positioned at the center of the semi-circular screen.
\update{Study results revealed no difference in pointing performance} between virtual and physical display platforms when users are positioned at the center of the semi-circular displays. 
In our second study, we explore the effect of users' location (relative to the center of the display) on pointing performance for both platforms. 
\update{Results showed no difference in users' pointing performance for the virtual and physical platforms for the investigated locations.}
Additionally, we observed that interaction distance (i.e., distance from the screen center to the user) significantly affects user performance.
Pointing from the center of the semi-circular display leads to higher performance and user preference. 

The contributions of the paper are:
\begin{itemize}
\item A comparison between two curved display platforms, i.e., virtual and physical, on users pointing performance with Fitts law
\item An exploration of the effect of user position relative to display on their pointing performance with  virtual and physical displays
\end{itemize}

\section{Background and Related Work}
We review works on users' interaction with physical and virtual environments and pointing tasks in physical and virtual displays.

\subsection{Comparison between Physical and Virtual Environments} 
Sharples et al. \cite{sharples2008virtual} examined the VR-induced sickness symptoms for four different displays (i.e., head-mounted display: Virtual research V8 resolution 1920$\times$480 FOV 60$\degree$, desktop (17" 800$\times$600), projection screen, and curved large displays [screen resolution 1024$\times$3556]) when viewing and interacting in a Virtual Factory (i.e., a program where the participants were instructed on potential health and safety hazards).
They found significantly higher levels of nausea, oculomotor, and disorientation in the Head-mounted display compared to the large curved display. 
Deisinger et al.\cite{deisinger1997effect} asked participants to identify boxes in their correct order on three different display platforms: HMD (240$\times$120), desktop screen (21" 1280$\times$1024), and wall-sized projection-based large display (144" 4096$\times$768). 
They found that participants experienced higher levels of immersion on the large wall-sized display compared to the two other platforms.  
Although previous literature found differences in subjective feedback between physical and virtual displays, they did not explore pointing performance (e.g., completion time and error rate for pointing) between virtual and physical displays.

\subsection{Large Physical Displays: Flat and Curved}
Prior work largely focused on pointing performance on large flat displays. 
Very few dealt with pointing performance on curved displays. 
However, we found no research on pointing performance on virtual curved displays. 
Research on large curved displays can be divided into two major streams: their comparison to flat displays, and the exploration of curved display specificities.

\textbf{Flat vs. Curved Large Displays:}
Shupp et al. \cite{shupp2009shaping} compared flat and curved large displays  for search, comparison, and overview tasks.
They found that curved displays outperformed flat displays for search (i.e., locating a symbol on a map) and comparison (finding differences between 2D graphs ) tasks. 
The study also shows that objects on a curved display are uniformly located from the user and thus are free from user bias (different regions do not suffer from performance issues). 
However, users have to rotate their heads more while using curved displays compared to flat ones.
Flat displays had an advantage over curved displays on overview tasks (i.e., state observations from visual data). 
However, users have to walk more for flat displays and suffer from region bias, where the task performance degrades as the objects are placed further away from the display center. 
A recent study by Liu et al. \cite{liu2020design} ran user studies to compare three different layouts: flat, quarter-circular, semi-circular, and full-circular for data visualization in a VR environment based on chart comparison and finding max value tasks. 
They found that for lower scale (12 visualizations placed on 3 $\times$ 4 grid) format, the flat layout outperformed curved layouts. 
However, when they increased the scale (36 visualizations placed on 12 $\times$ 3 grid) they found no difference in performance between the flat and curved layout. 
For the large-scale layout, the users preferred the half-circle layout compared to the flat layout as they had to walk less. 
Overall, studies have shown several advantages of curved displays with semi-circular layout over traditional flat ones; hence, we use a curved display in our study.

\textbf{Curved Displays Specificities:}
As compared to flat displays, curved displays have been shown to be more ergonomically suited for visual tasks for large-scale displays.
Kyung and Park \cite{kyung2021curved} compared 33" and 50" displays with different display curvatures (i.e., 40 cm, 60 cm, 120 cm radius and Flat) and recommended using a 60 cm radius for both displays for visual search tasks. 
Their result also found that with increasing size, curved displays do not face performance issues, unlike Flat displays. 
Nevertheless, there is an absence of research examining pointing tasks on large curved displays.
The sole study focusing on pointing on curved displays was a Fitts law study by Hennecke et al.\cite{hennecke2013investigating}, who used a 42" curved display to examine 2D pointing tasks using two inputs: direct touch and mouse inputs.
For both types of inputs, Fitts law was applicable. 
Due to technical limitations, the display was created from two HD projector outputs, and thus, was neither totally curved nor large in scale.
Ens et al. \cite{ens2016moving} conducted a Fitts law study in which the targets were arranged in a circular orientation using projectors in a CAVE environment.
 They considered five different field-of-view conditions (ranging from 8\degree to 128\degree). They observed improved pointing performance with a higher field-of-view; however, the improvement was not statistically significant.
In conclusion, pointing performance on large curved screens has been largely ignored.
Moreover, the research on curved displays did not take into consideration aspects specific to curved displays, such as interaction distance relative to display curvature.
Particularly, pointing performance research has only been done for physical displays, thus there is scope for advancement in this field with virtual ones.

\subsection{Target Pointing in HMD}
Researchers examined Fitts law tasks in virtual head-mounted displays.
Shi et al. \cite{shi2022pointing} conducted a Fitts law study in a head-mounted VR setting to determine the impact of various factors (i.e., input type, target width, and user distance)  on pointing performance.  
Their research revealed that the input mode, user distance, and target width all had significant effects on the movement time.  
Hansen et al. \cite{hansen2018fitts} compared three input types (i.e., eye gaze, head pointing, and mouse ) for pointing in HMDs and found that the mouse had the highest performance while head pointing was better than eye gaze.
Contrarily, eye gaze had a significantly lower subjective workload than head pointing.  
Batmaz et al. \cite{batmaz2019head} compared pointing performance in augmented and virtual reality using a pointing task that involved moving a spatially-tracked wand between front-back (i.e., viewing direction movement) and left-right(i.e., lateral movement). 
They demonstrated that movement direction (i.e., lateral and viewing direction) and target width had significant effects on movement time.  
Furthermore, for the viewing direction movements, front-to-back movements were significantly faster than back-to-front movements. 
Contrary to the previous claims \cite{jones2008effects, naceri2010depth}, they found no significant differences in pointing performance between augmented reality and virtual reality headsets. 
The aforementioned studies have only considered pointing performance in an HMD-based 3D environment.
However, researchers are yet to explore pointing performance on virtual interactable displays inside VR.

\subsection{Summary}
In contrast to the significant research \cite{kopper2010human, shoemaker2012two, janzen2016modeling} done on pointing performances on large flat displays, users pointing performance on large curved displays has remained unexplored.
However, for large displays, curved screens were preferred over flat displays \cite{liu2020design, kyung2021curved}.
Investigating and designing varying display properties of large curved displays are subject to hardware and financial restrictions\cite{cavallo2019dataspace}. 
Consequently, there is a lack of research that has investigated curve display-specific features, such as display curvature and off-centered interaction position \cite{Ullah2023LargeCurved, shupp2009shaping, kyung2021curved}.
Virtual displays inside VR has the potential to substitute physical displays in a variety of scenarios (e.g., for visual analytic tasks \cite{Ullah2023LargeCurved, cao2019large}). 
\update{Yet, prior work on HMDs primarily focused on pointing in 3D spaces with targets on flat planes, overlooking virtual displays and targets on curved surfaces. These curved surfaces can potentially impact users' pointing performance due to the characteristics of targets projected onto them, including factors such as users' position.} 
Therefore, a comparative study of user performance between physical and virtual \update{curved} displays is necessary to find the nuances between them.

\section{Design Space for Curved Displays Exploration}
We begin with a summary of the different variations of Fitts law investigated in prior work.
We next discuss the factors that we consider in a Fitts law to compare pointing performances between physical and virtual large curved displays.
Standard pointing variables (such as target size and amplitude) and curved-display specific variables (such as relative user position relative to display curvature) are both taken into account for our study.

\subsection{Fitts Law}
Fitts law is widely used for modelling pointing performance of 1D \cite{fitts1954information, card1978evaluation, jota2010comparisonGIpaper}, 2D \cite{mackenzie1992fitts, hoffmann1995effective, hoffmann2011performance, nancel2013high} and 3D \cite{triantafyllidis2021challenges, hoffmann2011performance, kulik2020motor, leusmann2021literature} pointing tasks. 
The Shannon-Fitts formula proposed by MacKenzie \cite{mackenzie1992fitts} is the most popular version of Fitts law (ISO 9241-9 \cite{iso20009241}) for pointing tasks. 
In their work, they calculated the  movement time for pointing via: 
\begin{equation} \label{eqn_Fittslaw_popular}
MT=a+b \log _{2}\left(\frac{A}{W}+1\right)
\end{equation}
Where A denotes the target amplitude, W denotes the target width, and a and b are empirically calculated constants.
The Index of Difficulty (ID) of a pointing task is equal to $\log _{2}\left(\frac{A}{W}+1\right)$ which represents the task
difficulty, i.e., the higher the ID value, the more challenging the task.
Various works have proposed Fitts law extensions by incorporating target height \cite{crossman1956measurement, mackenzie1992extending, hoffmann1994effect}, angular measurements \cite{kondraske1994angular, kopper2010human, kopper2011understanding}, and depth factors \cite{grossman2004pointing}. 
In these extensions, the original Fitts law was modified to include additional variables or adapt the law to specific contexts or devices. 

\subsection{Target Width, Target Amplitude, and IDs}
Prior work on target pointing mostly considered linear (e.g., in meters or pixels) target width and amplitude \cite{ fitts1992information, tao2021freehand, babic2018pocket6, lischke2016magic, liu2015gunslinger, haque2015myopoint, siddhpuria2018pointing, nancel2013high}. 
Later, angular measurements (i.e., in degrees or radians) of target width and amplitude were explored for pointing tasks  \cite{kondraske1994angular, kopper2010human, kopper2011understanding}. 
As we are using one display curvature for both displays, the Index of difficulty ($\log _{2}\left(\frac{A}{W}+1\right)$) values for both angular and linear measurements will be the same.
Therefore, we consider linear measurements for amplitude and width in our study. 
ISO 9241-9 \cite{iso20009241} standard recommends covering a wide range of IDs (between 2 bits and 8 bits) \cite{soukoreff2004towards,iso20009241} to represent the task difficulty - which we used in our studies.

\subsection{Interaction Distance}  
Interaction distance is defined by the distance from the user perpendicular to the center of the display \cite{tao2021freehand, janzen2016modeling, shi2022pointing}.
Prior works reveal that users' pointing performance varies based on the interaction distance \cite{kovacs2008perceptual, hourcade2012small, kopper2010human}. 
Researchers showed that it has a direct correlation with the target width in the display space, which further impacts the target selection time \cite{kovacs2008perceptual, hourcade2012small}: the further away from a display, the smaller the target.
To alleviate this confounding effect, Kopper et al. \cite{kopper2010human} integrated the interaction distance by using angular measurements of target width and amplitude. 
They showed that angular measures are more accurate for distal pointing tasks than linear measures.
Tao et al. \cite{tao2021freehand} evaluated the effects of interaction distance on pointing performance where they found that linear interaction distance significantly affected pointing accuracy and workload.
However, using angular measures is not yet commonly accepted in the HCI community \cite{shoemaker2012two, janzen2016modeling}.

\subsection{Position Relative to Screen Center}  
To our knowledge, no prior work considered pointing at objects while varying user positions from the display center for curved displays.
With flat displays, the user's position impacts the viewing angle, which impacts performances \cite{hourcade2012small}.
With curved displays, the position might have an impact on performance.
For instance, users at the center of the curve interact with a display uniformly distributed around them.
However, users with an offset from the center interact with screen portions closer than others.
In addition, limited FoV on HMDs might create differences in users' pointing performance based on users' position and head rotation to find the targets in virtual displays.
Thus, it warrants further investigation into how users' pointing performance varies between physical and virtual displays based on their position relative to the screen.

\subsection{Exploration Summary}

Based on the design space, we selected the following four critical factors to explore - target width, target amplitude, interaction distance, and user offset. 
We report on two user studies:
\begin{itemize}
    \item \textbf{Study 1: Compare physical and virtual large display from the center of the displays}: The first study aims to compare users' pointing performances between a physical and a virtual large curved display with a Fitts law. More specifically, we investigate whether there are differences in users' pointing performance between the two platforms, especially when users are located at the center of the curved displays. 
    
    \item \textbf{Study 2: Explore Physical and Virtual Curved displays from different User distance/offset}: In our second study, we explore the effect of factors specific to curved displays on users' pointing performance. More specifically, we investigated how users' pointing performance varies in two platforms when they are placed with an offset from the center of the screen.
        
\end{itemize}

\subsection{Goals and Hypotheses}
Our study has two main goals (G1, and G2):
\begin{itemize}
    \item G1) To identify differences in pointing performance between virtual and physical large displays from the center of the displays.
    \item G2) To evaluate the effect of user positions relative to the display on their pointing performance on both large curved display platforms.

\end{itemize}

To reach these goals, we formulate the following Hypotheses:
\begin{itemize}
    \item H1) There are no differences in users' pointing performance irrespective of display type i.e., physical or virtual from the center of the display.
    Indeed, as a human motor behavior model, Fitts Law should not reveal any discrepancies if users' location is the same across conditions (\textbf{G1}).
    \item H2) Users pointing performance between two platforms will not vary even if users are positioned with an offset from the center of the display (\textbf{G2}). 
    \item H3) Pointing from farther away from the screen will lead to faster selections as increasing interaction distance allows the users to acquire targets with lower hand and head movements (\textbf{G2}).

\end{itemize}



\section{User Study 1}

We first conduct a 1D Fitts law study comparing users pointing performance between virtual and physical displays. 
Prior work mostly explored users' interaction with curved displays while positioning  them at the center of the display \cite{shupp2009shaping, liu2020design}.
Therefore, in this study, we positioned users at the center of the semi-circular large curved display (in our case, 3.27m from the display). 

\begin{figure*}[h]
	\centering
	\includegraphics[width=2.0\columnwidth]{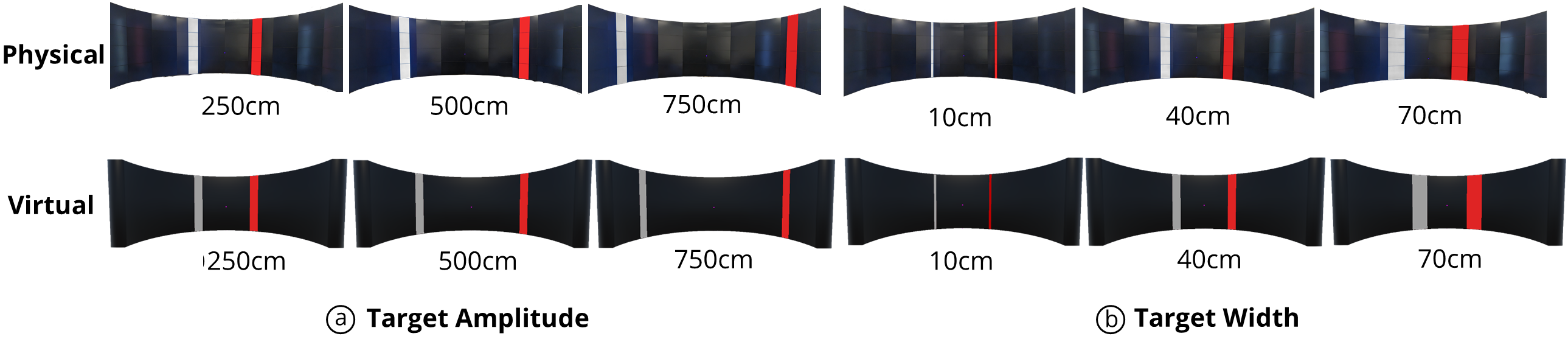}
	\caption{(a) Target Amplitude and (b) Target Width for physical and virtual displays}
	\label{fig:display-amplitude-width}
\end{figure*}

\subsection{Apparatus}
We used a physical curved display manufactured by Mechdyne \cite{mechdyne-new1} for this study, 
The display has forty 46-inch LED monitors arranged in 4 rows $\times$ 10 columns (\update{a portion of the display shown in Figure \ref{fig:teaser}b}). 
The 3D horizontally curved display has a 180\degree viewing angle.
The height of the display is 3 meters, and the radius of the semi-circular display is 3.27 meters.
The display is powered by CPU: Intel Xeon W-3224 8C/16T (3.5-4.6 GHz), RAM: 192GB DDR4 2933MHz (12 $\times$ 16GB), GPU: 4 X Nvidia RTX6000 with 1 x Nvidia Quadro SYNC II.
We used Unity 3D \cite{unity} to write applications to run on large physical displays.
To create the virtual display, we used the Oculus Quest 2 \cite{meta}, which has a horizontal field-of-view of 90\degree. 
We created a semi-circular display in the Unity game environment \cite{unity} with the same display height and radius as the physical setup.
To ensure a fair pointing comparison between the two platforms, we used the right controller of the Quest 2 with the raycasting technique \cite{takashina2021evaluation} for pointing at the displays. 
We placed markers on the Quest 2 controller and tracked its position and orientation using an 8-camera Optitrack System \cite{optitrack}. 
The Firebase Realtime Database \cite{google} was used for logging data during the study.  
Figure\ref{fig:teaser} a and b show both displays.


\subsection{Target Amplitude and Target Width}
We consider linear measurements \cite{shoemaker2012two, janzen2016modeling, mackenzie1992extending, soukoreff2004towards} for our study. 
We used 250cm, 500cm, and 750cm for the amplitude of our study (Figure \ref{fig:display-amplitude-width}a).
For the target width, we used 10cm, 40cm, and 70cm (Figure \ref{fig:display-amplitude-width}b). 
Once they were projected on the physical display, we ensured the linear measurements were correct with a measuring tape. 
In addition, for the VR application, the target widths and amplitudes were measured from the Unity application. 
Further, we used the pass-through feature of the Oculus Quest 2 \cite{meta} to overlay the virtual targets on top of the physical targets to avoid any discrepancies between the two platforms.
For instance, the 10cm target width and 250cm target amplitude on the virtual display perfectly overlapped with the 10cm target width and 250cm target amplitude on the physical display ensuring a 1-to-1 mapping.




\subsection{Participants}
We recruited 14 participants (7 male, 7 female; avg. 25.43, SD. 4.72) from our university. 
We used online advertisements as well as paper posters to recruit participants.
All of the participants were right-handed.
\update{Participants had an average of 1.56 years (SD. 2.02) of experience with VR HMDs and 0.14 years (SD. 0.36) of experience with large-scale displays. Note that participants reported the time of their first-hand experience when using VR to reflect their year of experience. Consequently, the reported values indicated a higher average experience with VR HMDs than with large-curved displays, as VR technology is more prevalent in the market. Only two individuals had no prior experience using a VR headset, while another two had experience with a large-scale display.}
Each participant received \$15 as compensation for participating in the study.


\subsection{Procedure}
A 1D Fitts law task was used in this study, where users were presented with two vertical bars representing targets.
For both platforms, the participants were placed at the center of the semi-circular display, which is 3.27m away from the screen.
Participants had to use the Quest 2 controller that was tracked with an OptiTrack System. 
From the controller, a ray was extended in the orthogonal direction to point at the large curved display (Figure \ref{fig:teaser}a and b).
Participants could move the ray in any direction by changing the orientation of the controller. 
Once they moved the ray tip (i.e., the end point of the ray) on top of the targets, they could press the trigger button to confirm a selection.   
Initially, we placed a `start' target at the center of the display for both conditions.
Once participants selected the `start' target, it disappeared and the application showed two new targets: colored red and grey.
They were to select the red-colored target, which switches between the two targets after each selection.
A selection outside the target region was logged as an error and re-queued among the unfinished trials.
Audio feedback was used to indicate successful and unsuccessful selections.

\begin{figure*}[htbp]
	\centering
	\includegraphics[width=2\columnwidth]{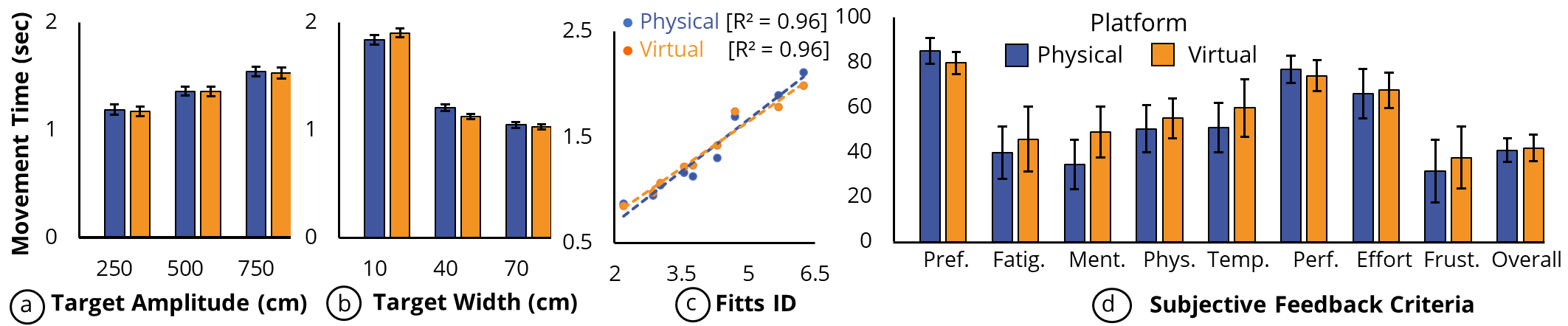}
	\caption{ Movement time for each platform by (a) target amplitude, (b) target width, (c) Fitts law Index of Difficulty, and (d) Subjective feedback (out of 100). Error bars represent 95\% Confidence Interval}
	\label{fig:exp1-all} 
\end{figure*}

\subsection{Design}
This study used a within-subject design with three independent variables: \textit{Platform} (physical and virtual), \textit{Target Amplitude} (250cm, 500cm, and 750cm) and \textit{Target Width} (10cm, 40cm, and 70cm).
The \textit{Target Width} $\times$ \textit{Target Amplitude} combinations lead to \textit{Index of Difficulty (ID)} ranging from 2.85 to 6.25 bits.
We applied a Latin square design to counterbalance the presentation order of the \textit{Platform} among the participants. 
Each \textit{Index of Difficulty (ID)} was repeated 10 times.
\textit{Target Amplitude} and \textit{Target Width} were randomly presented to participants.
Each participant had to perform a total of \update{180} selections (2 \textit{Platform} $\times$ 3 \textit{Target Amplitude} $\times$ \update{3} \textit{Target Width} $\times$ 10 repetitions). 
Additionally, the participants were allowed to take breaks to reduce fatigue.
Note that participants were provided with 45 practice trials before beginning with a \textit{Platform} to help familiarize themselves with the study conditions (i.e., 9 \textit{Index of Difficulty (ID)} $\times$ 5 repetitions).
Each participant took approximately 40 minutes to complete the experiment.

\subsection{Measurements}
Before starting the user study, participants were asked to provide demographic information. 
We recorded the movement time and error rate for each trial.
The movement time for a trial starts immediately after the previous selection and ends when participants press the trigger button to select the current target.
A trial is marked as an error if participants make a selection outside the target boundary.  
We used  the NASA-TLX \cite{nasatlx} to collect participants' perceived workload for the two platforms (physical and virtual).

\section{Study 1 Results}
\update{In both studies, Shapiro-Wilk tests were used to verify the normal distribution of the data.}
We analyze movement time and error with RM-ANOVA, and pairwise comparisons with Bonferroni corrections.
In case of sphericity violation, we report Greenhouse-Geisser corrected p-values and degrees of freedom.

\subsection{Movement Time}
 Figure \ref{fig:exp1-all}a-b show the Movement Time (MT) for different \textit{Target Amplitude} and \textit{Target Width} across two platforms, respectively. 
In terms of movement time, we did not find any significant effect of \textit{Platform} ($F_{1,13} = 0.17$, $p =0.69$) on movement time. 
The mean movement time is  1.37s CI: [1.35, 1.38] for the Virtual and 1.36s CI: [1.34, 1.38] for the Physical \textit{Platform}.
As expected, we found significant main effects for the independent variables \textit{Target Amplitude} ($F_{2,26}=110.72$, $p<0.001$, $\eta^2$=0.89), and \textit{Target Width} ($F_{2,26}=326.51$, $p<0.001$, $\eta^2$=0.96). 
For \textit{Target Amplitude}, the mean movement time is 1.18s (95\% confidence interval, CI: [1.08, 1.28]) for 250cm, 1.36s (CI: [1.27, 1.45]) for 500cm, and 1.54s (CI: [1.44, 1.64]) for 750cm (Figure \ref{fig:exp1-all}a). 
Post-hoc pairwise comparisons revealed that lower \textit{Target Amplitude} had significantly faster movement time than higher \textit{Target Amplitude} ($MT_{A=250cm}<MT_{A=500cm}<MT_{A=750cm}$, all $p<0.001$). 
For \textit{Target Width}, the mean movement time is 1.87s (CI: [1.79, 1.95]) for 10cm, 1.17s (CI: [1.12, 1.22]) for 40cm, 1.04s (CI: [0.99, 1.09])  for 70cm (Figure \ref{fig:exp1-all}b). 
Post-hoc comparison indicated that larger targets had significantly lower movement time than smaller targets ($MT_{W=70cm}<MT_{W=40cm}<MT_{W=10cm}$, all $p<0.001$).
We found significant interactions of \textit{Target Amplitude} $\times$ \textit{Target Width} ($F_{4,52}=14.46$, $p<0.001$, $\eta^2 = 0.53$) and \textit{Platform} $\times$ \textit{Target Width} ($F_{1.3,16.92}=4.64$, $p<0.05$, $\eta^2=0.26$).

\subsection{Error Rate}
There is no significant difference in the error rate for the different display \textit{Platform} ($F_{1, 13} = 0.64$, $p=0.43$). 
We found significant main effect of \textit{Target Width} ($F_{2,26}=30.84$, $p<0.001$, $\eta^2$=0.30) and \textit{Target Amplitude} ($F_{2,26}=7.12$, $p<0.01$, $\eta^2$=0.09) on error rate.
The smallest target (\textit{Target Width}=10cm, mean: 12.93\%, CI: [9.70, 16.15]) has a significantly higher error rate ($p<0.001$) than the other two \textit{Target Widths} (40cm, mean: 3.21\%, CI: [1.65, 4.76]; 70cm, mean: 2.18\%, CI: [0.72, 3.64]).
Targets placed the farthest apart (\textit{Target Amplitude}=750, mean: 8.86\%, CI: [6.04, 11.68]) have significantly higher error rates ($p<0.001$) than targets which are closest to the display center (\textit{Target Amplitude}=250, mean: 3.17\%, CI: [1.23, 5.11]).
Pairwise comparisons reveal no other significant differences.

\subsection{Fitts Law Regression Lines}
Figure \ref{fig:exp1-all}c shows the regression analysis that represents an average movement time for display \textit{Platforms} based on \textit{Index of Difficulty (ID)}. 
Overall, we found that the Fitts law model (Eq. \ref{eqn_Fittslaw_popular}) provides a fit of $R^2=0.96$ for both the physical and the virtual displays. 
This also indicates that the model could accurately explain about 96\% of the data points for both display platforms.
Interestingly, predictions have the same slope (b $\approx$ 0.3), indicating that both physical ($y=0.29x+0.19$) and virtual ($y=0.33x+0.04$) displays exhibit the same tolerance to an increase in task difficulty.
Note that physical and virtual platforms lead to very similar regression lines - which indicates that pointing tasks across these display platforms indeed lead to similar movement times. 
This further establishes that both display platforms have similar pointing performance.

\subsection{Preference Score}
We used NASA TLX \cite{nasatlx} questionnaire to collect users' feedback on their effort, mental demand, physical demand, temporal demand, performance, frustration and overall task load for \textit{Platform}. 
We also used another questionnaire to collect participants' general preference and fatigue for the physical and the virtual displays (Figure \ref{fig:exp1-all}d).
A Friedman test showed that pointing on the virtual display (mean: 48.92, CI: [37.60, 60.26]) is significantly ($\chi^2 (1, N=14)=9.00, p<0.01$) more mentally demanding than pointing on the physical one (mean: 34.28, CI: [23.26, 45.31]). 
There were no other statistically significant differences observed for other data.

\subsection{Discussion Study 1}
\subsubsection{Pointing performance comparison}
\textbf{H1:} no difference in pointing performance for the physical and virtual display is valid.
We found no significant difference between the two displays based on users pointing performance. 
This similarity in performance holds across different target widths and amplitudes.
The Fitts law further validates no difference in \update{users' pointing} performance between the two display platforms. 

\subsubsection{Subjective feedback comparison}
\update{The subjective feedback between the virtual and physical display showed no significant difference except for the mental demand criteria. Three participants mentioned that tasks were more cognitively demanding with VR compared to the larger display due to the limited Field-of-View (FOV) available in VR. This restriction in FOV posed greater mental demand for visually searching the targets. Based on participant feedback, we attribute this increased mental demand to the field-of-view difference between the two platforms.}
\section{User Study 2}

In our first study, we did not find any difference in users' pointing performance between the two platforms (virtual and physical) when participants are located at the center of the semi-circular screen.
Researchers found that users' pointing performance changes depending on users' position relative to the screen \cite{tao2021freehand, hourcade2012small, kovacs2008perceptual}.
Thus, our second study examines how users' pointing performance is affected based on their relative position from the center of virtual and physical displays.

\subsection{Interaction Distance \& User Offset}
\begin{figure}[htbp]
	\centering
	\includegraphics[width=1.0\columnwidth]{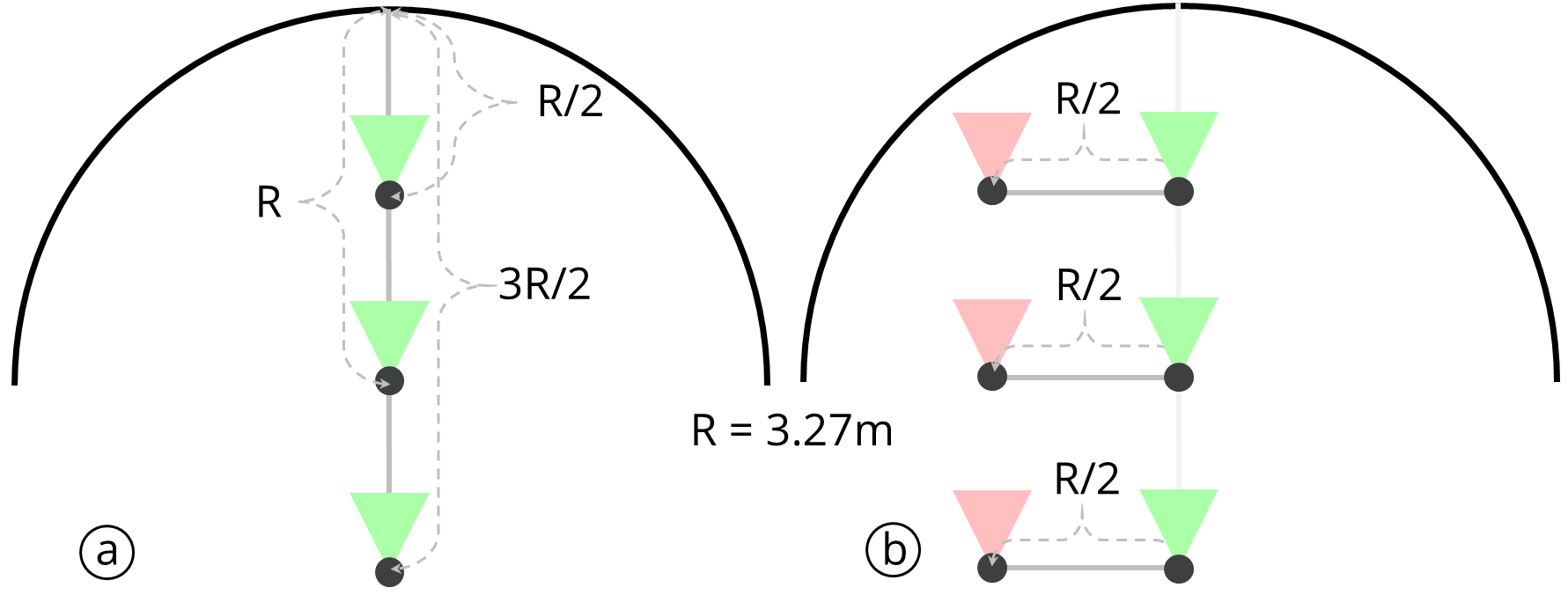}
	\caption{(a) Interaction Distance and (b) User Offset}
	\label{fig:interaction-offset} 
\end{figure}

We used the \textit{Interaction distance} as the distance from the screen on the \textit{main axis} (i.e., the axis that connects the middle of the screen and the center of the semi-circular display). 
In addition, we represent \textit{User offset} as the lateral displacement of the users' position from the `main axis'.
Figure \ref{fig:interaction-offset} shows the interaction distance and user offset from the center of the display.
While research has shown that interaction distance affects pointing performance for smaller displays \cite{tao2021freehand, hourcade2012small, kovacs2008perceptual}, we are unaware of any prior work that specifically addressed both of these relationships for wall-sized curved screens.
In addition, the offset is directly associated with the users' Field of View (FoV) in the VR condition. 
Therefore, it is critical to investigate the impact of the offset and compare it to the physical display condition to reveal the potential effects and implications of the FoV on user performance.


We investigate three interaction distances which are denoted by \(\frac{1}{2}\)R, R, and \(\frac{3}{2}\)R (shown in Figure \ref{fig:interaction-offset}a) where R is the radius of the semi-circular display (3.27m).
To illustrate, for our 3270R display - which has a radius of 3.27m, we consider the interaction distance R which is 3.27m, \(\frac{1}{2}\)R, which is 1.64m, and \(\frac{3}{2}\)R which is 4.91m from the center of the screen. 
A user can be positioned to the left or right of the main axis (i.e., \textit{User Offset}). 
Since target positions on axis-symmetric directions have no difference in pointing performance \cite{zhang2012extending,radwin1990method}, we place users on the left side of the main axis. 
We consider two offsets: No-offset (i.e., no lateral displacement from the main axis) and Left-Offset (c of \(\frac{1}{2}\)R from the main axis) as shown in Figure \ref{fig:interaction-offset}b. 
Thus, the \textit{Interaction Distance} and the \textit{User Offset} result in 6 different positions for a display.
We did this as in a real scenario interactive items and UI elements would be displayed on the curved display irrespective of the user's position in the VR environment.
We keep the target widths and amplitude uniform across trials, with linear measurements from the curve center to maintain consistency between different user positions.

\subsection{Participants and Apparatus}

We recruited 12 participants (6 male and 6 female; mean age 25.92 years, SD. 4.91) from the local University. 
\update{The participants' average prior experience was 1.57 years (SD. 2.19) for VR  HMDs and 0.18 years (SD. 0.37) for the large-scale display. Only three people had no experience using a VR headset, while only four had prior experience using large-scale displays.}
Each participant received a \$15 honorarium for their participation.

The participants had to interact will the same physical display (3270R large 3D curved large display) and virtual display (3270R rendered using the Oculus Quest 2).
To keep the input device consistent for both platforms, we used the same Quest 2 right controller with the raycasting technique for pointing at the displays. 
The movement of the controller was tracked with the Optitrack System \cite{optitrack}. 
A modified version of the two separate unity applications from Study 1 was used for logging user data.    

\subsection{Procedure \& Design} 
In Study 2, the participants had to perform the identical 1D pointing task as in Study 1, but from varying user positions (marked with ground indications illustrated in Figure \ref{fig:interaction-offset}).
We had five independent factors: \textit{Platform} (physical and virtual), \textit{Amplitude} (250cm, 500cm, and 750cm), \textit{Width} (10cm and 70cm), \textit{User Offset} (No-offset, Left), and the \textit{Interaction Distance} (\(\frac{1}{2}\)R, R, and \(\frac{3}{2}\)R). 
However, we were more interested in exploring the effect of the \textit{Platform}, \textit{User Offset}, and \textit{Interaction Distance} - which we considered as our primary factors for study 2.  
Based on the \textit{Amplitude} $\times$ \textit{Width} configuration, we had six IDs (ranging between 2.85 to 6.25 bits).
The participants had to repeat each task condition 10 times, resulting in 720 trials (2 \textit{Platform} $\times$  2 \textit{User Offset} $\times$ 3 \textit{Interaction Distance} $\times$ 3 \textit{Amplitude} $\times$ 2 \textit{Width} $\times$ 10 repetitions).
Trials are divided into 12 blocks where each block contains 10 repetitions of 6 IDs for each \textit{Platform}, \textit{Interaction Distance}, and \textit{User Offset}. 
After completing each block, the participants rested for two minutes to lower fatigue and motion sickness.
We applied a Latin square to counterbalance the order of presentation of the platform while randomizing other factors (i.e., \textit{Interaction Distance}, \textit{User Offset}, and IDs). 
Participants provided subjective feedback on each combination during the break. 
Note that participants had to complete at least five practice trials before starting a new block. 
The total time of the study was around 80 minutes per participant.



\begin{figure*}[htbp]
	\centering
	\includegraphics[width=1.8\columnwidth]{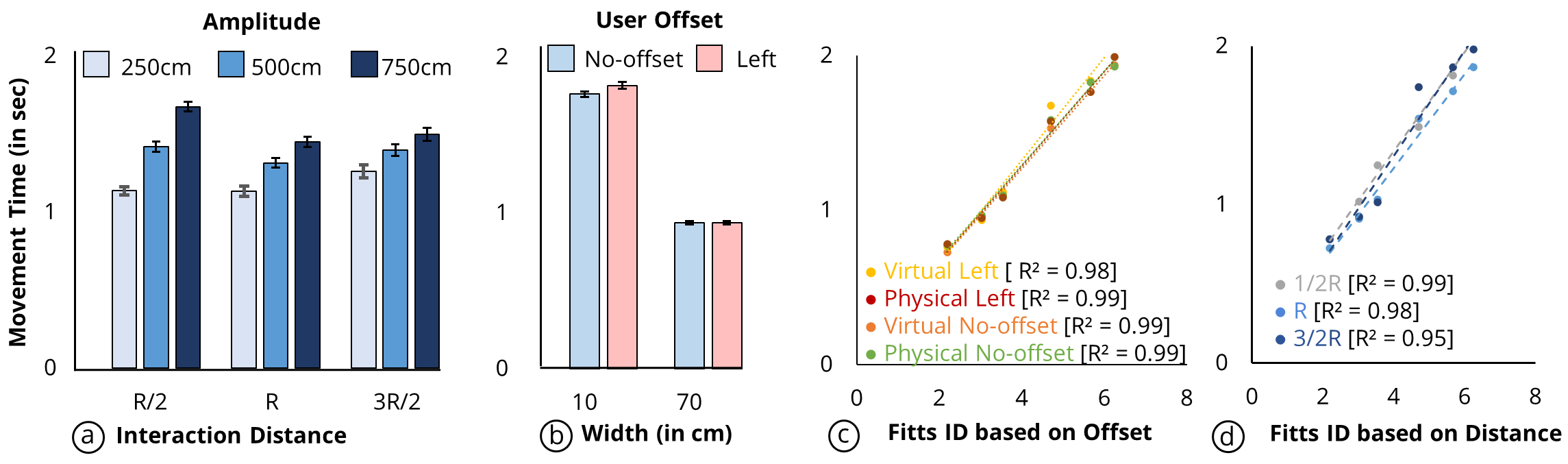}
	\caption{Movement Time by (a) Interaction distance for each target amplitude, (b) target width for each offset. Error bars represent 95\% CI. Fitts law model regression line for (c) user offset and (d) interaction distance}
	\label{fig:exp2-movementTime} 
\end{figure*}

\subsection{Measurement}
For each trial, movement time and error rate were recorded as we did in Study 1.
Demographic information and subjective feedback for each user offset and interaction distance condition was also collected.

\section{Study 2 Results}
We analyze movement time and error data with repeated measures ANOVA on log-transformed data and post-hoc pairwise comparisons with Bonferroni corrections.
In sphericity violation, we report Greenhouse-Geisser corrected p-values and degrees of freedom.

\subsection{Movement Time}

Interestingly, we again did not find any significant effect for \textit{Platform} on movement time ($F_{1,11}=0.02$, $p=0.89$). 
The mean movement time is 1.37s CI: [1.35, 1.38] for the Virtual and 1.36s CI: [1.34, 1.38] for the Physical \textit{Platform}. 
We found significant main effects of \textit{Amplitude} ($F_{2,22}=457.72$, $p < 0.0001$, $\eta^2$ = 0.98), \textit{Width} ($F_{1, 11}=252.78$, $p < 0.0001$, $\eta^2$ = 0.96), and \textit{Interaction Distance} ($F_{2,22}=11.63$, $p<0.001$, $\eta^2$ = 0.51) on movement time. 
The mean movement time is 1.41s (CI: [1.35, 1.47]) for \textit{Interaction Distance} \(\frac{1}{2}\)R, 1.3s (CI: [1.24, 1.35]) for R and 1.38s (CI: [1.32, 1.45]) for \(\frac{3}{2}\)R.
Post-hoc pairwise comparisons reveal that pointing at R is significantly faster than \(\frac{1}{2}\)R and \(\frac{3}{2}\)R (all $p<0.01$). 
In addition, pointing at \(\frac{3}{2}\)R is significantly faster than \(\frac{1}{2}\)R ($p<0.001$). 
For \textit{User Offset} ($F_{1,11}=0.77$, $p=0.40$), we  did not observe any significant effect on movement time.

We found a significant effect of \textit{Amplitude} $\times$ \textit{Interaction Distance} ($F_{4,44}=39.2$, $p<0.001$, $\eta^2 = 0.78$) in Figure \ref{fig:exp2-movementTime}a, \textit{Width} $\times$ \textit{Interaction Distance} ($F_{1.34,14.76}=54.5$, $p<0.001$, $\eta^2 = 0.83$) and \textit{Width} $\times$ \textit{Offset} ($F_{1,11}=4.92$, $p<0.05$, $\eta^2 = 0.31$) \ref{fig:exp2-movementTime}b.
While offset did not have an effect on selection times, for Width = 10cm, the selection was faster in the center (mean: 1.76s, CI:[1.72,1.8]) compared to the left position (mean: 1.82s, CI:[1.77,1.87]) (p<0.05).


\subsection{Error Rate}

We found no significant difference in the error rate for the different display \textit{Platforms} ($F_{1,11}=0.59$, $p=0.44$).
There is a significant difference in the error rate for \textit{Interaction Distance} ($F_{2,22}=14.31$, $p<0.001$, $\eta^2=0.04$) and \textit{Width} ($F_{1,11}=256.10$, $p<0.001$).
For \textit{Interaction Distance}, we found an error rate of (6.15\%, CI: [5.07, 7.24]) for \(\frac{1}{2}\)R, (8.33\%, CI: [7.10, 9.56]) for the R, and (10.30\%, CI: [8.97, 11.64]) for \(\frac{3}{2}\)R.. Post-hoc tests with Bonferroni corrections on \textit{Interaction Distance} showed that \(\frac{1}{2}\)R  has significantly lower rates (all $p < 0.01$) than R, and \(\frac{3}{2}\)R.
For \textit{Offset}, we found an error rate of 7.74\% (CI: [6.75, 8.73) for the no-offset position and 8.88\% (CI: [7.84, 9.91]) for the Left offset position. 
However, there is no significant difference in the error rate for \textit{Offset} ($F_{1,11}=3.08$, $p=0.08$) and \textit{Amplitude} ($F_{2,22}=3.08$, $p=0.11$).
Overall, we found Study 2 has higher average error rates as only two target widths were considered - leading to an increased proportion of 10cm \textit{Widths}.

\subsection{Fitts' Law Regression Line} 
The Fitts law equation for \textit{Offset} and \textit{Platforms} were $y=0.33x-0.01$ for Virtual Left, $y=0.31x+0.05$ for Physical Left, $y=0.30+0.05$ for Virtual No-offset, and $y=0.30x+0.08$. There was so little difference in movement time for \textbf{Offset} and \textit{Platform} that  the regression lines were indistinguishable ($R^2$ (all $\ge0.98$), Figure \ref{fig:exp2-movementTime}c). 
Additionally, we observe that the Fitts law could accurately model pointing performance($all R^2\ge0.95$ Figure) with varying \textit{Interaction Distance} (regression equations were \(\frac{3}{2}\)R: $y=0.33x-0.01$; R: $y=0.30x+0.05$; \(\frac{1}{2}\)R: $y=0.31x+0.09$).
We observed that pointing from interaction distance R is faster than the other two positions (3/2R and 1/2R) in Figure \ref{fig:exp2-movementTime}d.


\subsection{Preference Scores}
We collected users' feedback on the seven Nasa TLX criteria along with their preference, fatigue for \textit{Interaction Distance} and \textit{Offset}.

For \textit{Offset}, Friedman tests show that participants preferred being in the No-offset position compared to the Left for all of the 9 criteria [Preference ($\chi^2 (1,N=24)=6.54, p<0.05$), Fatigue ($\chi^2 (1,N=24)=6.32, p<0.05$), Mental Demand ($\chi^2 (1,N=24)=6.37, p<0.05$), Physical Demand ($\chi^2 (1,N=24)=8.05, p<0.01$), Temporal Demand ($\chi^2 (1,N=24)=9.94, p<0.01$), Performance ($\chi^2 (1,N=24)=8.03, p<0.01$), Effort ($\chi^2 (1,N=24)=6.55, p<0.05$), Frustration ($\chi^2 (1,N=24)=10.89, p<0.001$), and Overall task load ($\chi^2 (1,N=24)=8.91, p<0.01$)].
 
For \textit{Interaction Distance}, Friedman tests reveal significant differences for all subjective criteria [Preference ($\chi^2 (2,N=24)=22.94, p<0.001$), Fatigue ($\chi^2 (2,N=24)=25.54, p<0.001$), Mental Demand ($\chi^2 (2,N=24)=27.58, p<0.001$), Physical Demand ($\chi^2 (2,N=24)=27.35, p<0.001$) , Temporal Demand ($\chi^2 (2,N=24)=21.76, p<0.001$), Performance ($\chi^2 (2,N=24)=23.48, p<0.001$), Effort ($\chi^2 (2,N=24)=22.05, p<0.001$), Frustration ($\chi^2 (2,N=24)=21.16, p<0.001$), and Overall task load ($\chi^2 (2,N=24)=20.24, p<0.001$)].
Post-hoc pairwise comparisons (Bonferroni: $\alpha$-levels from 0.05 to 0.016) reveal that  R  and \(\frac{3}{2}\)R are significantly preferred than \(\frac{1}{2}\)R for all subjective criteria (all p<0.01). 

\subsection{Discussion Study 2}

\subsubsection{Comparison between the two platforms}
From the results, we found that our H2 is valid. 
Similar to study 1, we again found no significant effect of \textit{Platform} on movement time and error rate. 
This further highlights that there is no difference in pointing performance between virtual and physical displays when using the same setup, e.g., the same controller, screen size, target amplitudes, and widths. 
Participants even commented \textit{"Pointing at targets were identical in both platforms."} [P7, male, 21 years old] 

\subsubsection{Effect of Interaction Distance}
Our H3 on improved pointing performance as the user is placed farther away is not valid. 
We observed a relationship between interaction distance and movement time: users perform faster selection when the interaction distance is equal to the radius of the curved display. 
As seen from Figure \ref{fig:exp2-movementTime}a, being close to the display (interaction distance R/2) is beneficial for small amplitude movements (A=250cm) whereas being farther away (interaction distance 3R/2) is advantageous for large amplitude movements (A=750cm). 
\update{However, this relationship needs further investigation with more interaction distances to confirm the scalability of our findings. }
Eight out of twelve participants complained that being positioned closer to the display required constantly rotating their head and body to perform the reciprocal selection task for movements with large amplitudes (such as A=500cm and A=750cm).
\update{This observation aligns with findings from prior research \cite{liu2022effects}.}
Furthermore, they commented being farther away increased the pointer speed, which was difficult to control for targets close to the display center (A=250cm).
Being at the center (interaction distance R) optimizes the trade-off between interaction distance vs. performance and allows the users to make faster selections than the other interaction distance (i.e., R/2 and 3R/2).
When the user is located at the center of the display, the distance from the user to the target is the same irrespective of the target amplitude. 
These allow the users to select the targets faster using muscle memory (as suggested by participants. One participant commented \textit{"Selection from the center is better than the left as we can leverage muscle memory to select targets that are symmetrically positioned on the left and right sides."} [P11, male. 23 years old]. 
This observation is consistent with previous work which suggests that being positioned at the center of the semi-circular displays improves user task performance \cite{park2019effects}.
Two participants preferred to be positioned farther away as this allowed them to select targets with minimal body movement and ensured a larger portion of the display was visible to the user for both platforms. 
One participant stated \textit{"Being located farther away provided higher peripheral vision, especially for the virtual display as it had a low FOV."}  [P12, female, 24 years old]. 
Thus, we observed better subjective ratings for R and 3R/2.


\subsubsection{Effect of User Offset}
We found no difference in pointing performance between the center (i.e., no-offset) and the left Figure \ref{fig:exp2-movementTime}b. 
However, the subjective ratings show that no-offset positions are significantly better than the left positions.
As the users performed reciprocal selection tasks, targets were equally spaced around the display center. Thus, they preferred to be in the center rather than the left position. 
Comments from the participants revealed that users have to do different hand movements to select targets when positioned towards the left. 
In contrast, users can select the targets doing uniform hand movements when positioned in the center, allowing them to easily select targets using muscle memory.  
However, this feeling is due to the abstract reciprocal task. It might be different if - imagine - we propose an asymmetric task (e.g., text on the last, large pics on the right)

\section{Design Implications, Limitations and Future work}
We interpret key insights from our results and discuss the design implications for large curved displays. 
We highlight limitations observed during the studies and discuss potential future works.

\subsection{Design Implications}

\subsubsection{Virtual prototypes for Physical Interactive surfaces}
Results from the Studies indicate no significant difference in pointing performance between the two platforms. 
Study 2 further shows that similar performance is maintained for near or farther user placements and off-centered positions. 
Based on these results, we suggest that virtual displays in VR can be \update{a potential alternative for evaluating pointing-related tasks for large physical displays - reducing development time and cost of researching on large curved displays.} 



\subsubsection{Interaction Distance}
Study 2 results show that pointing from the center of the curved displays (interaction distance R and no-offset) is more favorable to the users.
Furthermore, interaction distance R/2 is better for the targets closer to the display center (lower amplitude), whereas interaction distance 3R/2 is better for targets farther away from the display center (high amplitude). 
Based on the results, we suggest that the center of the circular display (interaction distance = R) is the optimal location for interaction with the display, irrespective of the target location on the display - \update{especially when using a large curved display for pointing tasks.  
Also, note that the impact of interaction distance may vary depending on the display properties (e.g., display curvature) and specific tasks (e.g. map browsing).}

\subsubsection{User Offset}
\update{Based on the study results, we found no difference in pointing performance (movement time and error rate) for the center (i.e., no-offset) and the left position}, subjective feedback shows that the no-offset position is significantly better than the left position for all nine subjective feedback criteria. 
Therefore, we suggest placing users at the center of the displays with no offset on large curved displays. 

\subsection{Limitations and Future work}
The study was conducted between a physical and a virtual large display considering only one display radius of 3.27 meters (3270R) and a height of 3 meters. 
Future studies can evaluate whether these results are valid irrespective of display sizes (e.g., small, medium, and large displays) and display curvature (e.g., 2000R, 6000R). 
Due to the Oculus Quest 2 having a higher than usual Field-of-view (90\degree) and display resolution (1920 x 1832 for each eye), we did not notice any performance difference between the two display platforms. 
Future studies can evaluate how the results change when using different head-mounted displays with different FoV and display resolutions (e.g., Apple Vision Pro, Valve Index, HTC Vive, Quest 3, Microsoft Hololens).
Our study investigated users' pointing performance in a single-user scenario. 
Further research could evaluate whether users' pointing performance remains consistent between two platforms in collaborative multi-player scenarios.
\update{It is worth noting that our participant pool was relatively small, with 14 participants in Study 1 and 12 in Study 2. This sample size may not be representative of the broader population. Additionally, we intentionally excluded left-handed individuals, as our focus was solely on the left offset. In addition, incorporating left-handed participants would have necessitated examining both left and right offsets, complicating the study design. To improve the generalizability of our findings, future research should consider a larger and more diverse participant pool, encompassing a more comprehensive range of ages and handedness. }

\section{Conclusion}
In this paper, we investigated users pointing performance between a large curved physical and virtual display.
More specifically, we explored the effect of standard pointing factors (i.e., target width and amplitude) and curvature-specific factors (e.g., users' relative position to the display) on users' pointing performance between two displays.
We found comparable performance for both the virtual and physical on 1D pointing tasks irrespective of the investigated factors. 
We further showed that positioning users at the curvature center helps them to select targets efficiently compared to off-centered positions. 
\update{We provided guidelines for designers to leverage large virtual displays in VR for pointing-related tasks.} 





\bibliographystyle{ACM-Reference-Format.bst}
\bibliography{main}

\end{document}